\documentclass[12pt]{elsart}

\usepackage{epsfig,amsfonts}

\def\bbox#1{{ \mbox{\boldmath $#1$}} }

\def\Orth#1{\mathrm{O}(#1)}

\begin{document}

\begin{frontmatter}

\title{Finite size effects on  measures of critical exponents\\ 
in $d=3$ O($N$) models.}
\author{H.~G.~Ballesteros}{\footnote{\tt hector@lattice.fis.ucm.es}},
\author{L.~A.~Fern\'andez}{\footnote{\tt laf@lattice.fis.ucm.es}},
\author{V.~Mart\'{\i}n-Mayor}{\footnote{\tt victor@lattice.fis.ucm.es}},
\and
\author{A.~Mu\~noz Sudupe}{\footnote{\tt sudupe@lattice.fis.ucm.es}}.
\address{Departamento de F\'{\i}sica Te\'orica I, 
        Facultad de CC. F\'{\i}sicas,\\ 
\it Universidad Complutense de Madrid, 28040 Madrid, Spain.}

\maketitle

\begin{abstract}
We study the critical properties of three-dimensional $\Orth{N}$
models, for $N=2,3,4$.  Parameterizing the leading
corrections-to-scaling for the $\eta$ exponent, we obtain a reliable
infinite volume extrapolation, incompatible with previous Monte Carlo
values, but in agreement with $\epsilon$-expansions.  We also measure
the critical exponent related with the tensorial magnetization as well
as the $\nu$ exponents and critical couplings.
\end{abstract}

\end{frontmatter}

\section{Introduction}

The study of continuous spin models in low dimensions has been very
useful either for the knowledge of some physical systems directly
associated (mainly in condensed matter), or as toy systems for
studying relativistic field theories. 

The $\Orth{4}$ model in three dimensions has been conjectured to be 
on the same universality class as the finite-temperature chiral phase
transition of QCD with massless flavors~\cite{PISARSKY}. 
The $\Orth{4}$ universality
class has also appeared in perturbation-theory studies of spin-systems for
which the $\Orth{3}$ symmetry of the action is fully broken on the low
temperature phase~\cite{AZARIA1}.
It is well known that the
$\Orth{3}$ model in two dimensions offers a play-ground to explore
asymptotic freedom~\cite{POLYAKOV}, although there has been recently
some controversy on this point~\cite{O3SOKAL}.  Regarding
the applications to condensed matter physics, let us remind that the
three-dimensional $\Orth{3}$ model 
is the low-temperature's effective-model for a
bidimensional quantum antiferromagnet~\cite{O3COND}.  It also appears
as a limiting case of a $\mathrm{Z}_2$-gauge lattice model for nematic
phase transitions of liquid crystals~\cite{BETA2BETA}.  
Finally, the $\Orth{2}$ model in three dimensions is known to be in the
same universality class as superfluid $^4$He.

The $\Orth{N}$ nonlinear sigma models in  three dimensions 
have been extensively studied  either with
analytical~\cite{ZINNJUSTIN,ZINNJUSTINEPS} or numerical methods,
obtaining very accurate results in the determination of the critical
properties. In particular the critical exponents have been measured
for $N\le 4$ with a precision greater than a
1\%~\cite{O2,O3,O4}. However, the finite-size effects have not been
considered in a systematic way. We find this procedure not harmful
for $\nu$ exponent measures, but quite dangerous for $\eta$ exponents 
determinations.

Another point we focus on, is the study of the second-rank tensorial
magnetization. This composite operator is needed as an order-parameter
in some applications, for instance when the usual magnetization
vanishes, as in studies of nematics~\cite{BETA2BETA}.  It is also the
relevant order parameter for systems where the $\Orth{3}$ symmetry is
fully broken~\cite{AZARIA1}.  Naively one could expect that the
tensorial magnetization scales just as the square of the vectorial
one, that is, $\beta_\mathrm{T}$ is twice $\beta$, as a mean field
calculation predicts (it was also assumed in
ref.~\cite{BETA2BETA}). Nevertheless we shall show that this is not
so.  A somehow related question, which has aroused interest lately, is
the presence of some pathologies in the renormalization of composite
operators in $2+\epsilon$ expansions~\cite{WEGNER}.

In this paper we consider the $\Orth{2}$, $\Orth{3}$ and $\Orth{4}$
models, centering mainly in the measures of magnetic exponents using a
finite-size scaling method which is specially useful for observables
that change rapidly at the critical point. In addition to the standard
magnetic exponents we measure those related with tensorial
excitations. The  critical couplings and $\nu$ exponents are also studied. 
We are specially interested in the measure of $\nu$ in the $\Orth{4}$ case
in order to compare with the results obtained in the 
antiferromagnetic $\mathrm{RP}^2$ model~\cite{RP2LETTER}, which is related
with $\Orth{4}$ when there is a total breakdown of its $\Orth{3}$
symmetry.

\section{The model}

We consider the usual Hamiltonian
\begin{equation}
\mathcal{H}=-\beta \sum_{<i,j>}\bbox{v}_i\cdot\bbox{v}_j,
\end{equation}
where $\bbox{v}_i$ is a $N$ components normalized vector, and the
sum is extended over first neighbor pairs.

It is well known that this model undergoes a second order phase
transition for which the normalized magnetization
$\bbox{M}=\frac{1}{V}\sum_i \bbox{v}_i$ is an order parameter ($V$ is
the lattice volume). As this model can be simulated using cluster
algorithms~\cite{WOLFF}, it is possible to thermalize very large 
lattices.

We are also interested in studying the behavior of composite operators
that could be related with bound states from a quantum field theory
point of view. One can construct orthogonal states to that
generated by the fundamental field, just ensuring that the composite
operator transforms as a higher order irreducible representation. The
simplest representation beyond the fundamental one is the second rank
tensorial representation. The associated tensorial magnetization can
be written as
\begin{equation}
\mathcal{M}^{\alpha \beta}
=\frac{1}{V}\sum_i (v^\alpha_i v^\beta_i-\frac{1}{N}\delta^{\alpha\beta}).
\label{MTEN}
\end{equation}

As it happens with the vectorial magnetization, the mean value of
$\mathcal{M}$ is zero in a finite lattice, so, in the Monte Carlo
simulation we have to construct an estimator that avoids the
tunneling effects. We define the (normalized) magnetizations as  
\begin{equation}
M=\left\langle \sqrt{\bbox{M}^2}\right\rangle \quad ,
\quad M_\mathrm{T}=\left\langle \sqrt{\mathrm{tr} \mathcal{M}^2}\right\rangle.
\end{equation}

We also define the associated susceptibilities as
\begin{equation}
\chi=V \left\langle 
                \bbox{M}^2\right\rangle\quad ,\quad
\chi_\mathrm{T}=V \left\langle 
                \mathrm{tr}\mathcal{M}^2\right\rangle.
\end{equation}

The critical behavior of those quantities is expected to be

\begin{equation}
M\propto t^{\beta}\quad ,\quad 
\quad M_\mathrm{T}\propto t^{\beta_\mathrm{T}}\quad ,\quad 
\chi\propto |t|^{-\gamma}\quad ,\quad 
\chi_\mathrm{T}\propto |t|^{-\gamma_\mathrm{T}},
\end{equation}
where $t$ is the reduced temperature.
We expect the tensorial exponents to be also related through the
(hyper) scaling relation $\gamma_\mathrm{T}+2\beta_\mathrm{T}=\nu d$.

\section{The method}

To compute the critical exponents we have used a finite-size scaling
analysis. Specifically, we have used the method of
refs.~\cite{RP2LETTER} that consists in the comparison of 
observables on several pairs of lattices at the same coupling. The mean
value for the operator $O$,  measured in
a length $L$ lattice, at a coupling $\beta$ in the critical region, is
expected to behave as

\begin{equation}
\langle O(L,\beta) \rangle=L^{x_O/\nu}\left[F_O(\xi(L,\beta)/L) + 
L^{-\omega}G_O(\xi(L,\beta)/L) + \ldots\right],
\label{FSS}
\end{equation}

where $F_O$ and $G_O$ are smooth scaling functions for this operator,
$\xi$ is a measure of the correlation length and $x_O$ is the critical
exponent associated with $O$. Let us recall that $\omega$ is an
universal exponent related with the first irrelevant operator. The
dots stand for further scaling corrections. We have also dropped a
$\xi^{-\omega}$ term, negligible in the critical region. To eliminate
the scaling functions we compare the measures in two different lattice
sizes ($L_1,L_2$) at the same coupling.  Let be
\begin{equation}
Q_O=\frac{\langle O(L_2,\beta)\rangle}{\langle O(L_1,\beta)\rangle}=
s^{x_O/\nu}\frac{F_O\left(\xi(L_2,\beta)/L_2\right)}
       {F_O\left(\xi(L_1,\beta)/L_1\right)}+O(L^{-\omega}),
\end{equation}
where $s=L_2/L_1$. It is easy to find the coupling value where
$Q_\xi=s$. Measuring $O$ at that point we obtain the critical
exponent from
\begin{equation}
\left.Q_O\right|_{Q_\xi=s}=s^{x_O/\nu}+O(L^{-\omega}).
\label{QO}
\end{equation}

We remark that even if $O$ is a fast varying function
of the coupling in the critical region, as the magnetization, the
statistical correlation between $Q_O$ and $Q_\xi$ allows a very 
precise determination of the exponent.

For the correlation length, we use a second momentum
definition~\cite{LONGCOR} which 
is easy to measure and permits to obtain an accurate value: 
\begin{equation}
\xi=\left(\frac{\chi/F-1}{4\sin^2(\pi/L)}\right)^{1/2},
\label{XI}
\end{equation}
where $F$ is defined as the Fourier transform of the two-point correlation 
function at minimal momentum ($(2\pi/L,0,0)$ and permutations).

\section{Critical exponents}

For the Monte Carlo simulation, we have used the Wolff's embedding
algorithm with a single cluster update~\cite{WOLFF}. We have simulated
in lattice sizes from $L=8$ to $L=64$ on the critical coupling reported on
refs.~\cite{O2,O3,O4}. 
We have used the spectral density method to extrapolate 
to the neighborhood of these couplings.
We have updated 25 million clusters for O(2) and O(3) and 50 million
in the O(4) case. The autocorrelation times
are very small in all cases (not larger than a hundred of clusters). 
The runs have been performed on several workstations.

We have used the operator $\d \xi/\d \beta$ to obtain the $\nu$
exponent ($x_{\d \xi/\d \beta}=1+\nu$). 
For the magnetic exponents, we
use the total magnetization ($x_M=\beta$) as well as the corresponding
susceptibility ($x_\chi=\gamma$). From them we obtain the $\eta$
exponent using the scaling relations
$\gamma/\nu=2-\eta,2\beta/\nu=d-2+\eta$. For the tensorial channel we
obtain a different set of exponents that we denote as
$\beta_T,\gamma_T,\eta_T$ respectively.

In table~\ref{ONTAB} we report the results for the
exponents displaying also the used operator. We have checked  
other observables as well as other definitions of the 
correlation length but, in all cases, either the corrections-to-scaling or 
the statistical errors are greater. 

\begin{table}[t]
\caption{Critical exponents obtained from a finite-size scaling
analysis using data from lattices of sizes $L$ and $2L$ 
for the O($N$) models. In the second
row we show the operator used for each column.}
\smallskip
\begin{center}
\begin{tabular}{|r||r|l|l|l|l|l|}\hline
&   & \multicolumn{1}{c|}{$\nu$} 
    & \multicolumn{2}{c|}{$\eta$} 
    & \multicolumn{2}{c|}{$\eta_T$}      \\
\cline{3-4} \cline{4-5} \cline{6-7}
N &$L$ & \multicolumn{1}{|c|}{$d\xi/d\beta$} 
    & \multicolumn{1}{c|}{$\chi$}      
    & \multicolumn{1}{c|}{$M$} 
    & \multicolumn{1}{c|}{$\chi_T$}        
    & \multicolumn{1}{c|}{$M_T$}\\\hline\hline
2&8 & 0.683(3) &0.0252(10)&0.0297(11)&1.499(2)&1.506(2)\\\cline{2-7}
&12& 0.678(3) &0.0303(11)&0.0333(13)&1.496(3)&1.500(3)\\\cline{2-7}
&16& 0.672(3) &0.0329(12)&0.0355(12)&1.494(2)&1.497(3)\\\cline{2-7}
&24& 0.676(4) &0.0344(12)&0.0355(13)&1.494(3)&1.495(3)\\\cline{2-7}
&32& 0.670(3) &0.0366(12)&0.0387(13)&1.494(3)&1.496(3)\\\hline\hline

3&8 &0.724(3)&0.0300(10) &  0.0317(10) & 1.432(2)  & 1.437(2)\\\cline{2-7}
&12 &0.712(4)&0.0337(9)  &  0.0352(10) & 1.4312(17)& 1.4342(17)\\\cline{2-7}
&16 &0.712(4)&0.0344(11) &  0.0354(12) & 1.428(2)  & 1.429(2)\\\cline{2-7}
&24 &0.716(5)&0.0378(12) &  0.0385(13) & 1.4320(18)& 1.4335(18)\\\cline{2-7}
&32 &0.711(5)&0.0371(11) &  0.0377(12) & 1.428(2)  & 1.430(2)\\\hline\hline

4&8& 0.752(2) &0.0307(6)& 0.0316(6)&  1.3735(10)& 1.3767(11)\\\cline{2-7}
&12& 0.747(3) &0.0338(5)& 0.0345(6)&  1.3771(10)& 1.3790(10)\\\cline{2-7}
&16& 0.754(4) &0.0341(7)& 0.0344(7)&  1.3770(14)& 1.3781(13)\\\cline{2-7}
&24& 0.757(4) &0.0348(5)& 0.0349(5)&  1.3771(10)& 1.3774(11)\\\cline{2-7}
&32& 0.753(5) &0.0359(9)& 0.0361(10)&  1.3753(18)& 1.3759(18)\\\hline
\end{tabular}
\end{center}
\bigskip
\label{ONTAB}
\end{table}

\section{Infinite volume extrapolation}

In table~\ref{ONTAB} corrections-to-scaling are clearly 
visible for the $\eta$ exponent.
To control them, we need an estimation of the $\omega$ exponent. 
To measure this index and the critical coupling~\cite{RP2LETTER}, we study 
the crossing between
the Binder cumulant of the magnetization for different lattice
sizes, as well as the corresponding for $\xi/L$. 
The shift from the critical coupling of the crossing point for lattices 
of sizes $L_1$ and $L_2$ behaves as~\cite{BINDER}
\begin{equation}
\Delta\beta^{L_1,L_2}\propto
\frac{1-s^{-\omega}}{s^{\frac{1}{\nu}}-1}L_1^{-\omega-\frac{1}{\nu}}.
\label{SHIFTBETA}
\end{equation}
As the crossing point for the Binder cumulant and $\xi/L$
tends to the critical coupling from opposite sides, it is convenient to fit
all the data together.
We have carried out two types of fits, one fixing the smaller lattice
and the other fixing the $L_2/L_1$ ratio. In table~\ref{BETATAB} we present 
the results obtained using the full covariance matrix, for $L_1=8$ and $s=2$.

We observe a value for the $\omega$ exponent compatible with the
$0.78(2)$ value obtained from $g$-expansions for the
O(2) and O(3) models~\cite{ZINNJUSTIN}. In the O(4) case the result
is almost twice. Lacking a theoretical prediction, this could be 
interpreted in two ways, the exponent could truly be so big, or it might
be that the coefficient of the leading corrections-to-scaling term
is exceedingly small. 

\begin{table}[t]
\caption{Fits for $\beta_{\mathrm{c}}(\infty)$ and the corrections-to-scaling 
exponent $\omega$.  The second error bar in $\omega$ is due
to the variation of $\nu$ exponent within a 1\% interval (the size of
previously published error bars).}
\smallskip
\begin{center}
\begin{tabular}{|c|c|l|l|l|}\hline
N & Fit & \multicolumn{1}{c|}{$\chi^2/ \mathrm {d.o.f.}$} 
        & \multicolumn{1}{c|}{$\omega$} 
        & \multicolumn{1}{c|}{$\beta_{\mathrm{c}}(\infty)$}      \\
\cline{1-5}
2 &$L_1=8$  &11 /8  &0.86(12)(3)  &0.454169(4)\\\cline{2-5}
  &$s=2$    &9.8/6  &0.81(12)(1)  &0.454165(4)\\\hline  
3&$L_1=8$   &2.0/8  &0.64(13)(2)  &0.693001(10)\\\cline{2-5} 
 &$s=2$     &2.3/6  &0.71(15)(1)  &0.693002(12)\\\hline
4 &$L_1=8$  &11.9/8 &1.80(18)(6)  &0.935858(8)\\\cline{2-5} 
  &$s=2$     &8.0/6 &1.85(21)(2)  &0.935861(8)\\\hline
\end{tabular}
\end{center}
\label{BETATAB}
\medskip
\end{table}

The values obtained for the critical couplings are extremely precise,
and compatible with the most accurate previous determinations by 
Monte Carlo simulations~\cite{GOTTLOB,O3,O4}. Let us remark that the
conjectured value for the critical coupling of the $\Orth{3}$ 
model~\cite{LOG2}, $\beta_c=\log 2$, is ten standard deviations 
away from our measures.

To control finite-size effects the most common procedure is 
to pick the smaller lattice for which there is
no scaling corrections. That is, one uses a log-log plot, discard small
lattices, and stop when the value of critical index stabilizes. Using
this method, it has been determined that for the $\Orth{2}$
model it is enough to use lattices with sizes $L\geq 16$~\cite{O2},
$L\geq 12$ for the $\Orth{3}$ model~\cite{O3}, and $L\geq 8$ for the
$\Orth{4}$ model~\cite{O4}.
We shall confirm this assumption for the $\nu$
exponent but no for the magnetic ones.

It might not be possible to find a {\it safe} $L_\mathrm{min}$ lattice 
(in fact, to find scaling corrections is just a matter of 
statistical accuracy), we thus need an extrapolation procedure.   
With an $\omega$ estimation, we can extrapolate to the infinite volume
limit, with an ansatz for the exponent $x_O$:

\begin{equation}
\left.\frac{x_O}{\nu}\right|_{\infty} -\left.\frac{x_O}{\nu}
\right|_{(L,2L)}\propto L^{-\omega}.
\label{Xw}
\end{equation}

The situation is fairly different for the $\nu$ and $\eta$ exponents,
therefore we shall discuss them separately.

\subsection{$\eta$ type exponents.}

Due to the high accuracy that we get from the statistical correlation 
between the measures of $\xi$ and $\chi$, we are able to resolve 
finite-size corrections. One could wonder if only the first correction
term ($\propto L^{-\omega}$) is needed. To check it we have used an objective 
criterium:
 we perform the fit from $L_\mathrm{min}$, then we repeat it discarding 
$L_\mathrm{min}$ and check if both fitted parameters (slope and extrapolation) 
are compatible. If that is the case, we take the central value 
from the fit with 
$L_\mathrm{min}$ and the error bar from the fit without it. 
In the three models, we find that $L_\mathrm{min}=8$ is enough for this
purpose.

\begin{figure}[t]
\centering\epsfig{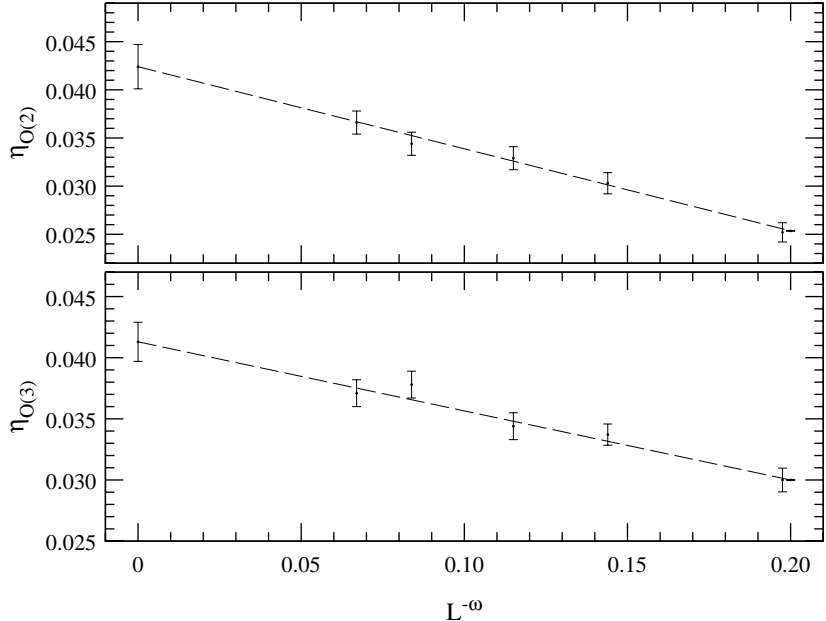}
\caption{
$\eta$ estimation from $\chi$ for pairs of lattices of sizes $L$ and 2$L$, as a function of $L^{-\omega}$, $\omega=0.78$, in the $\Orth{2}$ model (upper part),
and  $\Orth{3}$ model (lower part). Lines correspond to the $L_\mathrm{min}=12$
fit.
}
\label{ETA}
\end{figure}

For the $\Orth{2}$ and $\Orth{3}$ models, we have a very precise
knowledge of $\omega=0.78(2)$, from series analysis. We plot in
figure~\ref{ETA} our data as a function of $L^{-\omega}$. For
$\Orth{2}$, the fit has $\chi^2/\mathrm{d.o.f.}=0.85/3$.  For
$\Orth{3}$ we obtain $\chi^2/\mathrm{d.o.f.}=2.98/3$.  We then find no
reason to expect higher order corrections to be significant.  
We have also performed a simulation on a $L=6$ lattice, finding that, for  
the $\Orth{2}$ model, the corresponding $\eta$ value is one standard deviation 
away from the corresponding point in the fit performed for 
$L_\mathrm{min}=12$. Therefore, it seems even reasonable to keep the error from
the $L_\mathrm{min}=8$ fit ($\eta=0.0424(14)$).
 For the $\Orth{3}$ model, the $L=6$ point is three standard 
deviations away from the fit for $L_\mathrm{min}=12$ due to higher order
corrections-to-scaling. Notice in fig.~\ref{NU} that the slope for
$\Orth{2}$ is significantly larger, which could mask higher order corrections.
For the
$\Orth{4}$ model, lacking a theoretical knowledge of $\omega$, there
are stronger uncertainties. We find the change from $\omega=0.78(2)$
for $\Orth{2}$ and $\Orth{3}$, to $\omega=1.8(2)$ for $\Orth{4}$ very
surprising (see table~\ref{BETATAB}). This might arise from an
unexpected cancelation of first order scaling-corrections for the
Binder cumulant and the correlation length. We show in
table~\ref{ETAON} the extrapolated values for $\eta$ exponents. For
the $\Orth{4}$ model we present the extrapolation with $\omega=0.78$
($\chi^2/\mathrm{d.o.f.}=2.9/3$) and with $\omega=1.8(2)$ with
$\chi^2/\mathrm{d.o.f.}=1.0/3$. Both values are hardly compatible.

The determination of $\eta$ from $\chi$ and ${M}$ are of course
coincident in all cases due to the strong statistical correlation
between both observables. We find that the value for $\Orth{2}$ and
$\Orth{3}$ are compatible with $\epsilon$-expansions and not too far
from g-expansions, but incompatible with the previous Monte Carlo
results~\cite{O2,O3,O4}.  For $\Orth{4}$ both $\eta$ values are
significantly lower than for the other models which is not surprising
because, in the large $N$ limit, this exponent should go to zero.
 
\begin{table}[b]
\caption{Infinite volume extrapolation for $\eta$ from $\chi$ and ${M}$.
The second error is due to the uncertainty on  $\omega$. For the 
$\Orth{4}$ model, the first column has been calculated with $\omega=0.78$ and
the second with $\omega=1.8(2)$. In the last three rows we present, respectively, the results 
for this exponent from $\epsilon$, g-expansions and previously published
Monte Carlo values.}
\smallskip
\begin{center}
\begin{tabular}{|c|c|c|c|c|}\cline{1-5}
&\multicolumn{1}{c|}{$\Orth{2}$} 
        & \multicolumn{1}{c|}{$\Orth{3}$} 
        & \multicolumn{2}{c|}{$\Orth{4}$}      \\\cline{1-5}
$\chi$&0.0424(23)(2)&0.0413(15)(1)&0.0384(12)&0.0359(6)(3) \\\cline{1-5}
\cline{1-5}
${M}$&0.0421(25)(2)&0.0414(18)(1)&0.0381(13)&0.0359(6)(2) \\\cline{1-5}
\multicolumn{1}{|c|}{$\epsilon$-expansion}
&\multicolumn{1}{c|}{0.040(3)~\cite{ZINNJUSTIN}}
&\multicolumn{1}{c|}{0.040(3)~\cite{ZINNJUSTIN}}
&\multicolumn{2}{c|}{0.03(1)~\cite{O4EPSEXP}}\\\cline{1-5}
\multicolumn{1}{|c|}{g-expansion}
&\multicolumn{1}{c|}{0.033(4)~\cite{ZINNJUSTINEPS}}
&\multicolumn{1}{c|}{0.033(4)~\cite{ZINNJUSTINEPS}}
&\multicolumn{2}{c|}{--}\\\cline{1-5}
\multicolumn{1}{|c|}{Previous MC}
&\multicolumn{1}{c|}{0.024(6)~\cite{O2}}
&\multicolumn{1}{c|}{0.028(2)~\cite{O3}}
&\multicolumn{2}{c|}{0.025(4)~\cite{O4}}\\\cline{1-5}
\end{tabular}
\end{center}
\label{ETAON}
\end{table}

Let us consider the $\eta_T$ exponent.
From table~\ref{ONTAB} corrections-to-scaling are not self-evident.
If one adopts this optimistic point of view and performs a mean of all values 
in the table one finds that only for $L\geq 16$, $\chi^2/\mathrm{d.o.f.}$ 
becomes acceptable. 
Errors for the means are much smaller than for individual measures, 
and so, maybe smaller than scaling-corrections. 
We show these mean values on table~\ref{ETATONOMEGA}.
We think safer to perform a fit to the functional form~(\ref{Xw}).
We show the extrapolated values on table~\ref{ETATONOMEGA}. Errors coming
from the uncertainty on $\omega$ are smaller than a 10\% of the quoted errors.
Even more, the two values for $\Orth{4}$ are compatible.
The value of $\chi^2/\mathrm{d.o.f.}$ for $\Orth{4}$ and $\Orth{2}$ models is
about $0.6/3$. For the $\Orth{3}$ model is higher (about $6/3$).
Regarding the conjectured relation $2\beta=\beta_T$ proposed in 
ref.~\cite{BETA2BETA}, notice that it can be formulated  equivalently
as $1+2\eta=\eta_T$ which is absolutely ruled out by our data.

For the comparison between the critical exponents of the $\Orth{4}$ model
and those of the antiferromagnetic $\mathrm{RP}^2$ one, notice that
the $\eta$ value for $\Orth{4}$ is fully compatible with the corresponding 
exponent for the staggered magnetization of the $\mathrm{RP}^2$ model(
$\eta_\mathrm{stag}=0.0380(26)$~\cite{RP2LETTER}). The $\eta$ exponent 
associated with the usual magnetization for the $\mathrm{RP}^2$ model
($\eta_{\mathrm{RP}^2}=1.339(10)$~\cite{RP2LETTER}) is not compatible
with $\eta_T$, the difference being of order $\eta$.

\begin{table}[b]
\caption{For every model we show in the first column the mean values for
$\eta_T$ from $\chi_T$ and $M_T$ for $L\geq 16$. The second columns
display the infinite volume extrapolation for $\eta_T$ from $\chi_T$ and 
${M}_T$.For the 
$\Orth{4}$ model, the second column has been calculated with $\omega=0.78$ and
the third with $\omega=1.8(2)$.}
\smallskip
\begin{center}
\begin{tabular}{|l|c|c||c|c||c|c|c|}\cline{1-8}
&\multicolumn{2}{c||}{$\Orth{2}$} 
        & \multicolumn{2}{c||}{$\Orth{3}$} 
        & \multicolumn{3}{c|}{$\Orth{4}$}      \\\cline{1-8}
$\chi_T$ &1.494(1)&1.489(4)&1.431(1)&1.427(3)&1.3766(5)&1.374(5)
& 1.376(2)\\\cline{1-8}
${M}_T$&1.496(1)&1.491(4)&1.429(1)&1.427(3)&1.3773(5)&1.375(5)
& 1.376(2)\\\cline{1-8}
\end{tabular}
\end{center}
\label{ETATONOMEGA}
\end{table}

\subsection{$\nu$ exponent.}

\begin{figure}[t]
\centering\epsfig{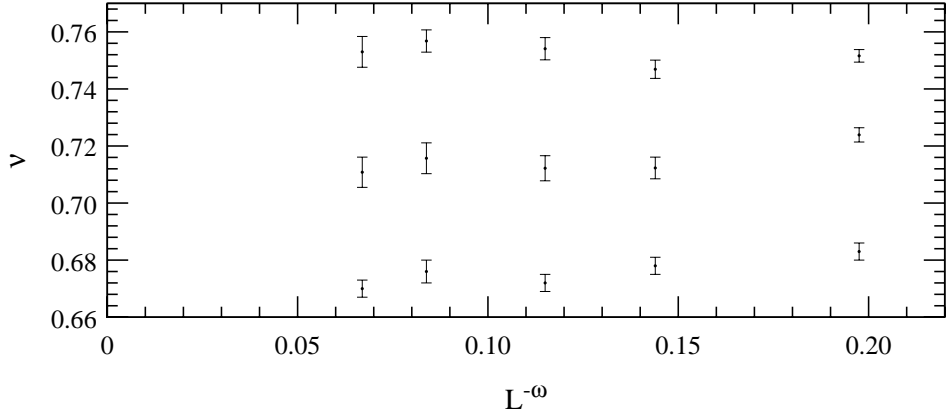}
\caption{
$\nu$ estimation from pairs of lattices of sizes $L$ and $2L$, as a
function of $L^{-0.78}$, in the $\Orth{4}$  (upper part),
$\Orth{3}$  (middle part), and  $\Orth{2}$ model (lower part).
}
\label{NU}
\end{figure}

In figure~\ref{NU}, we plot the estimation of $\nu$ that we get from
pairs of lattices of sizes $L$ and 2$L$ as a function of $L^{-\omega}$.
Indeed one would be tempted to claim that for the $\Orth{4}$ model, 
finite-size effects are beyond our resolution for $L\geq 8$, as stated in 
ref.~\cite{O4}. For the $\Orth{3}$ model this also seems to be the case
for $L\geq 12$~\cite{O3}. However the question is not so clear 
for the $\Orth{2}$ model.
Nevertheless, the experimental value is known to be
$\nu^{\mathrm{exp}}=0.6705(6)$~\cite{NU_O2}, and we do find  that 
for $L\geq 16$ our $\nu$ estimation is consistent with it.
If we actually believe that the corrections-to-scaling are negligible
for $L\geq L_{\mathrm{min}}$, we can take the mean of the {\it safe} lattices,
getting

\begin{equation}
\begin{array}{lclc}
\nu_{\Orth{2}} &=& 0.6721(13) &, \chi^2/\mathrm{d.o.f.}=1.7/2,\\
\nu_{\Orth{3}} &=& 0.7128(14) &, \chi^2/\mathrm{d.o.f.}=0.6/3,\\
\nu_{\Orth{4}} &=& 0.7525(10) &, \chi^2/\mathrm{d.o.f.}=3.4/4.
\end{array}
\end{equation}

Error bars decrease strongly compared to the data in table~\ref{ONTAB},
therefore it is not clear if scaling-corrections are still negligible.
A more conservative point of view would ask for the consideration of
these corrections. For this we use the ansatz~(\ref{Xw}).
However, from the plot in figure~\ref{NU}, it seems clear that the
scaling corrections for $L=8,12$ in the $\Orth{2}$ model, and
$L=8$ in the $\Orth{3}$ model, could hardly be linear on $L^{-\omega}$.
Performing the fit suggested by equation~(\ref{Xw}), with $\omega=0.78(2)$
for $\Orth{2}$ and $\Orth{3}$~\cite{ZINNJUSTIN}
from the {\it safe $L_\mathrm{min}$}, we get a compatible
value, but with a very increased error bar:

\begin{equation}
\begin{array}{lcl}
\nu_{\Orth{2}} &=& 0.670(10), \\
\nu_{\Orth{3}}&=& 0.711 (10).\\
\end{array}
\end{equation}

In the $\Orth{4}$ case, we perform the extrapolation with both $\omega$ values.
The extrapolation does not increase significantly the error bars.
Therefore we consider two values of $L_\mathrm{min}$. 
For $L_\mathrm{min}=8$ we obtain $\nu=0.754(3)$ for $\omega=0.78$  and
 $\nu=0.7531(15)(3)$ for $\omega=1.8(2)$. On the other hand, with 
 $L_\mathrm{min}=12$ we get $\nu=0.765(8)$
for $\omega=0.78$ and $\nu=0.7585(34)(7)$, for $\omega=1.8(2)$.

Regarding the relation with the antiferromagnetic $\mathrm{RP}^2$
model ($\nu=0.783(11)$), we find it unlikely that both values could
coincide, but we cannot rule it out. Notice that if we accept the value
$\omega=1.8(2)$  we would find a significant difference on this
universal exponent from $\mathrm{RP}^2$ ($\omega=0.85(5)$).

\section{Conclusions}

We have obtained accurate measures of critical exponents and couplings
for three dimensional O($N$) models. The method used, based on the
finite-size scaling ansatz, has a remarkable performance when
computing magnetic exponents.  The values for $\eta$ exponents
presented are incompatible with previous Monte Carlo results and they
have smaller statistical errors. We should point out that our
values agree with those obtained with $4-\epsilon$ expansions, and are
not incompatible with those computed by means of  $g$ expansions, 
in opposition with previous results. We also show that the statistical
accuracy that can be reached on three dimensional systems is such that
the strongest uncertainties come from finite-size effects, for which
a reliable theoretical parameterization would be highly desirable.

We present measures of the exponent corresponding to the tensorial
magnetization, which is not twice the corresponding to the usual
magnetization as previously stated. 

We increase significantly the precision of previous measures of the
exponents of O(4), showing that the differences with the $\nu$ exponent
from the values of the antiferromagnetic RP$^2$
model~\cite{RP2LETTER} are two standard deviations appart.
However whether they belong to the same universality class or not is 
not yet completely established. 

This work has been partially supported by CICyT AEN93-0604-C03-02 and
AEN95-1284-E.

\end{document}